\documentclass[conference,a4paper]{APSIPA2021}
\usepackage{multirow}
\usepackage{graphicx}
\usepackage{amsmath}
\usepackage{amssymb}
\usepackage{amsxtra}
\usepackage{threeparttable}
\usepackage{cite}

\renewcommand{\baselinestretch}{0.99}

\begin{document}

\title{Implementation of interactive tools for investigating fundamental frequency response of voiced sounds to auditory stimulation}

\author{%
\authorblockN{%
Hideki Kawahara\authorrefmark{1}, 
Toshie Matsui\authorrefmark{2}, 
Kohei Yatabe\authorrefmark{3}, 
Ken-Ichi Sakakibara\authorrefmark{4}, 
Minoru Tsuzaki\authorrefmark{5}, \\ 
Masanori Morise\authorrefmark{7}, and 
Toshio Irino\authorrefmark{1}
}
\authorblockA{%
\authorrefmark{1}
Wakayama University, Wakayama, Japan\hspace{0.3cm}
E-mail: \{kawahara,irino\}@wakayama-u.ac.jp}
\authorblockA{%
\hspace{-3.75cm}\authorrefmark{2}
Toyohashi University of Technology, Aichi, Japan\hspace{0.3cm}
E-mail: tmatsui@cs.tut.ac.jp}
\authorblockA{%
\hspace{-0.45cm}\authorrefmark{3}
Waseda University, Tokyo, Japan\hspace{0.3cm}
E-mail: k.yatabe@asagi.waseda.jp}
\authorblockA{%
\hspace{-2.9cm}\authorrefmark{4}
Health Science University of Hokkaido, Japan\hspace{0.3cm}
E-mail: kis@hoku-iryo-u.ac.jp}
\authorblockA{%
\hspace{-1.9cm}\authorrefmark{5}
Kyoto City University of Arts, Kyoto Japan\hspace{0.3cm}
E-mail: minoru.tsuzaki@kcua.ac.jp}
\authorblockA{%
\hspace{-0.9cm}\authorrefmark{7}
Meiji University, Tokyo, Japan\hspace{0.3cm}
E-mail: mmorise@meiji.ac.jp}
}

\maketitle
\thispagestyle{empty}

\begin{abstract}
We introduced a measurement procedure for the involuntary response of voice fundamental-frequency to frequency modulated auditory stimulation.
This involuntary response plays an essential role in voice fundamental frequency control while less investigated due to technical difficulties.
This article introduces an interactive and real-time tool for investigating this response and supporting tools adopting our new measurement method.
The method enables simultaneous measurement of multiple system properties based on a novel set of extended time-stretched pulses combined with orthogonalization.
We made MATLAB implementation of these tools available as an open-source repository.
This article also provides the detailed measurement procedure using the interactive tool followed by offline measurement tools for conducting subjective experiments and statistical analyses.
It also provides technical descriptions of constituent signal processing subsystems as appendices.
This application serves as an example for adopting our method to biological system analysis.

\end{abstract}

\section{Introduction}
Without feedback regulation, the fundamental frequency ($f_\mathrm{o}$\footnote{We use symbol $f_\mathrm{o}$ to represent the fundamental frequency adopting the discussion in the forum article\cite{titze2015jasaforum}.}) of sustained vowels cannot keep constant value\cite{titze1994book}.
Auditory feedback of speakers' voices plays an essential role in this regulation\cite{TOURVILLE20081429}.
This feedback regulation consists of voluntary, intentional control, and involuntary automatic control functions.
We recently found that the voice $f_\mathrm{o}$ systematically responds to auditory stimulation presented while voicing a sustained vowel keeping pitch constant.
Based on this finding, we proposed a procedure\cite{kawahara2021mixture} for investigating this involuntary control behavior by adopting a new simultaneous measurement method using extended time-stretched pulses\cite{kawahara2020simultaneous,kawahara2021icassp}.

The purpose of this article is complementary to our article, which focused on theoretical aspects of the method for investigating the response to auditory stimulation.
This article presents detailed procedures of typical experiments, and settings for other researchers can replicate our tests that measure involuntary voice $f_\mathrm{o}$ response to auditory stimulation.
We believe our procedure provides an objective method for investigating human pitch perception mechanisms, especially for the fine temporal structure of the auditory stimulation.
This involuntary response also provides a non-invasive method to insert perturbations into the voice production process.

\section{Background}
This section overlaps significantly with the background section of our article\cite{kawahara2021mixture} accepted for Interspeech2021.
We need this overlap to introduce our motivation and the experiment's goal, which our designed tool conducts.

Without proper regulation, we are not able to keep the fundamental frequency
of the voice (for example, sustained vowels) constant\cite{titze1994book}.
Auditory feedback plays an essential role in 
this regulation\cite{jones2008auditory,TOURVILLE20081429,houde2013PNAS}.
Vibrato, which makes singing voice attractive, also involves auditory feedback in production\cite{leydon2003role,titze2002reflex}.
Despite decades of research on voice fundamental frequency control mechanisms, it still is a hot topic\cite{larson2016sensory,behroozmand2020modulation,murray2020relationships,peng2021causal}.
Note that the target of the regulation is not the $f_\mathrm{o}$ value.
The target is the perceived pitch and is a psychological attribute,\cite{Moore2013book}.
For periodic signals, $f_\mathrm{o}$ value is the perceived pitch's physical correlate.
In other words, we can observe the perceptual attribute, pitch, directly using the $f_\mathrm{o}$ value
of the produced voice.
The regulation of voice pitch consists of voluntary and involuntary control\cite{hain2000instructing,zarate2010neural}.
The shifted pitch paradigm\cite{burnett1997voice} used in these studies has difficulty investigating this involuntary response.

The first author proposed to use a pseudo-random signal\cite{schroeder1979integrated} to perturb the
$f_\mathrm{o}$ of the fed-back voice.
It enabled to make the test signal unpredictable and to derive
the impulse response of the auditory-to-voice $f_\mathrm{o}$ chain\cite{kawahara1994interactions,kawahara1996icslp}.
This unpredictability enabled the measurement of involuntary response to pitch perturbation.
However, it was difficult for others to replicate the test
because it required a complex combination of hardware and
software tools.
The procedure also consisted of several drawbacks due to
available technology in the 1990s.
For example, we measured the response to pitch perturbation using the maximum length sequence (MLS)\cite{schroeder1979integrated}.
Selection of MLS among other TSP signals~\cite{aoshima1981jasa,dunn1993distortion,farina2000simultaneous,stan2002comparison,guidorzi2015impulse} was inevitable to
make the test signal unpredictable.
However, MLS has difficulty in measuring systems with non-linearity\cite{farina2000simultaneous,stan2002comparison}.
Conventional pitch extractors are the other source of the problem.
They introduced non-linear and unpredictable distortions in the
extracted $f_\mathrm{o}$ trajectories.

We succeeded in making test signals which are unpredictable and do not have MLS's difficulty.
Our new system analysis method uses a new extended TSP called CAPRICEP (Cascaded All-Pass filters with RandomIzed CEnter frequencies and Phase Polarities)\cite{kawahara2021icassp}.
We used CAPRICEP and developed an auditory-to-speech chain analysis system by adopting its simultaneous measurement of linear, non-linear, and random responses\cite{kawahara2020simultaneous}.
We developed an instantaneous frequency-based $f_\mathrm{o}$ analysis method instead of using conventional pitch extractors and removed the distortions\cite{kawahara2021mixture} mentioned above.
Combining these analysis methods and substantially advanced
computational power removed all the difficulties in measuring the auditory-to-speech chain response. Consequently, they resulted in an easy-to-use tool for conducting experiments\cite{kawahara2021mixture}. 

We speculate that this involuntary response measurement introduces a new strategy in pitch perception research.
Psychoacoustics and physiology are two prominent approachs to pitch perception\cite{carlyon1994jasa,cheveigne2005pitch,Moore2013book,lyon2017human}.
The tool provides access to the internal representation of ``pitch'' skipping cognitive and perceptual processes involved in psychoacoustic experiments.
The experiments using resolved and unresolved harmonic components and with and without fundamental component is an essential step.
The objective response analysis using test stimuli having different fine temporal structures, such as sine, cosine, and alternating phase setting of harmonic components\cite{patterson1987jasa} is a promising direction.
Including the low peak-factor signals\cite{schroeder1970synthesis} in this experiment is an informative extension.

Our article\cite{kawahara2021mixture} focused on theoretical aspects of the involuntary response measurement method.
The descriptions of the use and procedure of testing such responses were not enough for the readers to replicate the results.
We provide missing details of the tool and the test procedure in this article.
The tool is open-sourced and available from the first author's GitHub repository\cite{kawahara2020gitHk}.
We hope the combination of these makes the involuntary response measurement accessible to researchers.

\section{Experiment descriptions and procedures}
The goal of the experiment is to measure the involuntary $f_\mathrm{o}$ response to the auditory stimulation with $f_\mathrm{o}$ modulation.
The modulation is random to make it unpredictable for measuring the involuntary response.
We prepared different types of test signals for investigating pitch perception mechanisms.

\begin{figure}[tbp]
\begin{center}
\includegraphics[width=\hsize]{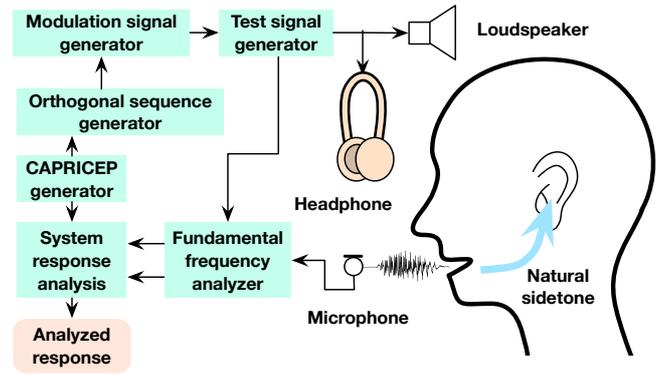}
\caption{Schematic diagram of experiments.}
\label{fig:spRespTestRev}
\end{center}
\end{figure}
Figure~\ref{fig:spRespTestRev} shows a schematic diagram of the experiment.
The subjects' task is to produce a sustained vowel keeping pitch constant while listening to a test sound.
The presentation of the test sound has two modes, and one mode uses a headphone the other mode uses a loudspeaker.
We developed an interactive and real-time tool for experimenting.
\begin{figure}[tbp]
\begin{center}
\includegraphics[width=\hsize]{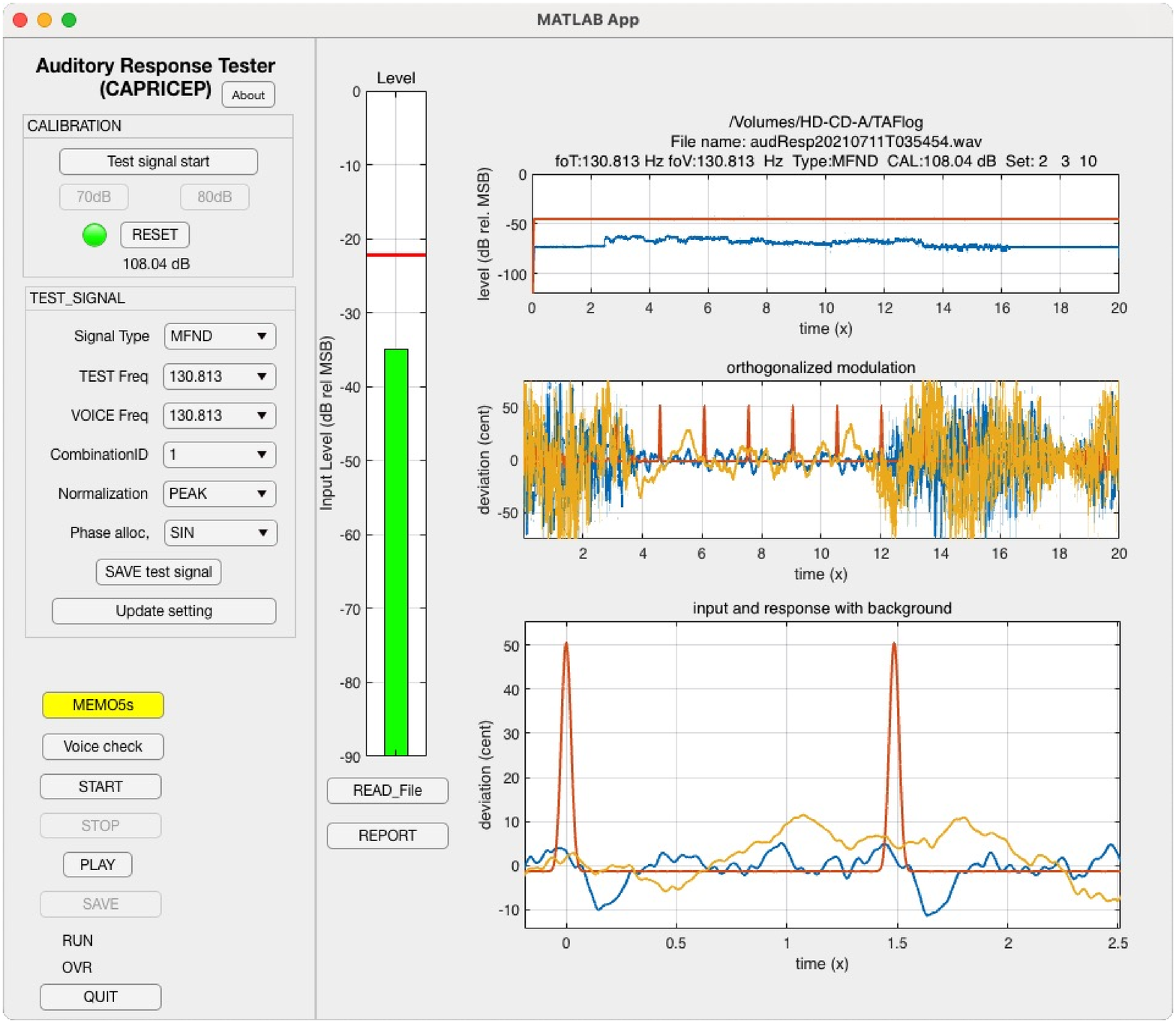}
\caption{GUI of the interactive test tool. The left panel is for control, and the right panel displays the analysis results and inspects saved test records.}
\label{fig:auditoryTestGUI}
\end{center}
\end{figure}

\subsection{Subjects' task}
We prepared two tasks, the same pitch task and a different pitch task.
1)~Same pitch task: We instructed the participant ``Please start voicing about one second after the test tone starts. Please use vowel \texttt{/X/} keeping pitch constant. Please use the same pitch of the test sound.''
2)~Different pitch task: We asked the participant to practice voicing using a target sound with a different pitch than the test sound. Then we instructed the participant ``Please start voicing about one second after the test sound starts. Please use vowel \texttt{/X/} keeping pitch constant. Please use the pitch you practiced. It is different from the test sound.''

\subsection{Test signal and target signal}
Figure~\ref{fig:auditoryTestGUI} shows the GUI of the tool which controls the experiment and analyses the result.
The left panel is for control, and the right panel is for displaying the analysis results.

\subsubsection{Signal type}
The left-center sub-panel controls the test signal settings.
The top pulldown menu determines the types of the test signal.
We prepared four types of test signals.
1)~The first is a frequency modulated sinusoid designated as ``\texttt{SINE}'' in the GUI.
2)~The second is a complex sound (sum of harmonically related frequency-modulated sinusoid) with the fundamental component and following 19 components, ``\texttt{SINES}.''
3)~The third is a complex sound without the fundamental component, in the other word missing-fundamental sound, ``\texttt{MFND}.''
4)~The fourth is a missing-fundamental sound consisting of only higher components, ``\texttt{MFNDH}.''
The following two pulldown menus define the $f_\mathrm{o}$ of the test signal and the target sound.

\subsubsection{Signal attributes}
The following ``\texttt{CombinationID}'' pulldown menu defines the combination of constituent three unit-CAPRICEPs.
We selected ten different unit-CAPRICEPs and used twenty sets from 720 possible combinations.

The following ``\texttt{Normalization}'' pulldown menu defines the level adjustment criteria.
We used four criteria.
1)~Peak level normalization: This sets the maximum instantaneous amplitude of the signal to 0.8 (1 is full scale), designated as ``\texttt{PEAK}.''
2)~RMS level normalization: This sets the RMS (Root Mean Squared) value of the signal -26~dB to the full scale, ``\texttt{TOTAL\_RMS}.''
3)~Component level normalization: This sets the amplitude of the fundamental component to -30~dB to the full scale, ``\texttt{COMPONENT}.''

The following ``\texttt{Phase alloc.}'' pulldown menu defines the phase relation between components.
For multiple component signals, we prepared four options of their phase relations, a sine phase ``\texttt{SIN},'' a cosine phase ``\texttt{COS},'' an alternating phase (sine and cosine for every other component) ``\texttt{ALT},'' and the Schr{\"o}der phase ``\texttt{SCH}.''

\subsubsection{Frequency modulation}
We used a mixture of orthogonalized sequences made from extended time-stretched pulses (CAPRICEP: Cascaded All-Pass filters with RandomIzed CEnter frequencies and Phase polarity), processed by a pink noise shaper, for modulating the $f_\mathrm{o}$\cite{kawahara2021mixture}.

\subsubsection{Definition of settings and saving test signals}
The following two buttons provide means to save the generates test signals and updating menu items and test signal details.
The ``\texttt{Save test signal}'' button saves the generated test signal using the wave format with metadata representing test signal attributes.
The  ``\texttt{Update setting}'' button updates menu items and the frequency modulation depth of the test signal by reading an experiment condition definition file.

\subsection{Procedure}
This section describes a few examples of the procedures consisting of test sessions.
It starts with the preparation of the test environment and calibration of the input system.

\subsubsection{Test environment}
We implemented the GUI tool and other analysis tools using MATLAB with Signal Processing Toolbox and Audio Toolbox.
We used \texttt{appdesigner} for developing the GUI tool.
Table~\ref{tbl:equipment} shows a typical environment for conducting experiments described in this article.
\begin{table}[tbp]
\caption{Typical environment required for conducting the tests described in this article.}
\begin{center}
\begin{tabular}{c|c}
\hline
Equipment & Description \\ \hline \hline
Computer system & macOS ,  Windows 10 \\ \hline
Software & MATLAB  \\ 
  (Toolbox)       &  Signal Processing and Audio Toolbox \\ \hline
Audio Interface & 441000~Hz sampling and 16~bit or more \\ \hline
Microphone & Omni directional or cardioid pattern \\ \hline
Headphone & Circumaural and/or noise cancelling \\ \hline
Loudspeaker & Usable from 100~Hz to 10000~Hz \\ \hline
Sound level meter & IEC Class-2 or better \\ \hline
\end{tabular}
\end{center}
\label{tbl:equipment}
\end{table}%

\subsubsection{Start up procedure}
When the tool starts, it asks the location of the storage.
Then, it asks to select the audio interface (combined with the driver), which can perform simultaneous input and output%
\footnote{For macOS, it also asks to select the input audio interface and the output audio interface. The experimenter needs to select the same hardware. This additional procedure is a walk-around for the performance issue of macOS real-time audio processing.}.

\subsubsection{Calibration}
For making the acquired voiced sounds, it is better to adopt the recommendations\cite{Rita2018ajsp,svec2010ajsp}.
It is essential to calibrate the sensitivity of the acquisition system to calculate the sound pressure level of the produced voice.
The top left sub-panel is for this calibration.

First, set the sensitivity of the audio interface to the highest sensitivity while preventing overloading.
Ask the participant to produce the loudest voice for this adjustment.
Then, click the ``\texttt{Test signal start}'' button to start playback the pink noise.
Adjust the audio interface's output (and the amplifier) to make the sound pressure level (measured using A-weighting) at the microphone position 70~dB or 80~dB.
When the level is appropriate and settled, click the relevant (70~dB or 80~dB) button.
The green light turns on, indicating that the system is calibrated. The following text shows the calibration information (necessary gain to convert the acquired value to the sound pressure level).

After calibration, do not change the sensitivity of the input system.
When adjustment is unavoidable, use the ``\texttt{Reset}'' button to re-calibrate.

\subsubsection{Pre-test vocalization check}
Before starting a test, the experimenter may use the ``\texttt{Voice Check}'' button to let the participant try voicing at the target pitch.
The click of the button starts the target sound, which is the same type as the test signal without frequency modulation.
\begin{figure}[tbp]
\begin{center}
\includegraphics[width=\hsize]{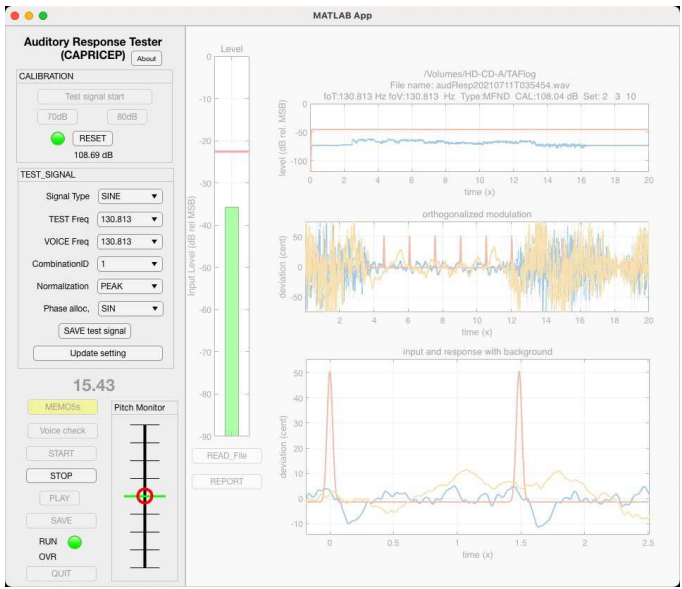}
\caption{Snapshot while checking voice pitch control. The pitch indicator appears at the bottom part of the left panel. The green horizontal line indicates the target pitch location. The red circle represents the $f_\mathrm{o}$ of the acquired signal in real-time. The interval of the horizontal lines is one semi-tone.}
\label{fig:voiceCheck}
\end{center}
\end{figure}

The experimenter uses this pitch monitor to instruct the participant how to adjust voice $f_\mathrm{o}$ to the target pitch.
Note that participants without some musical training sometimes cannot understand instructions such as ``please make the pitch higher'' and ``please lower the pitch.''

\subsubsection{Test vocalization}
After completing the voice check, the test starts by clinking the ``\texttt{Start}'' button.
As mentioned in the task description, the experimenter gives one of the following instructions to the participant.

1)~Same pitch task: We instructed the participant ``Please start voicing about one second after the test tone starts. Please use vowel /X/, keeping pitch constant. Please use the same pitch of the test sound.''
2)~Different pitch task: We asked the participant to practice voicing using a target sound with a different pitch than the test sound. Then we instructed the participant ``Please start voicing about one second after the test sound starts. Please use vowel /X/, keeping pitch constant. Please use the pitch you practiced. It is different from the test sound.''

We usually use /a/ for the vowel /X/.
The test signal duration is 20 seconds.
It is challenging to sustain voicing this long.
It is usable when the length of voicing exceeds 10 seconds.
Please record the distance between the participant's mouth and the microphone to make the acquired data re-usable\cite{Rita2018ajsp}.

While the test signal is on, the timer displays the time from the start, just under the test signal sub-panel.
After completion of the test signal, the ``\texttt{PLAY}'' button and the ``\texttt{SAVE}'' button go to active. 
The experimenter can check the recording by clicking the ``\texttt{PLAY}'' button for glitches and trouble in the recording.
If everything is relevant, clicking the ``\texttt{SAVE}'' button saves the acquired data using a time-stamp-based unique name.

\subsubsection{Analysis and logging}
When the ``\texttt{SAVE}'' button saves the acquired data, the analysis procedure starts using the saved data.
The right panel of the GUI displays the analysis results using three graphs.
Note that the experimenter cannot see the analysis results before saving the acquired data.
This experiment pipeline design eliminates the chance of cherry-picking in the data acquisition.

The top two graphs are for troubleshooting afterward.
The bottom graph shows the response to the auditory stimulation.
The horizontal axis represents the time from the maximum stimulation.
The vertical axis represents the magnitude of the stimulation and the response.
The unit is musical cent.
The red line shows the stimulation, and the blue line shows the response.
The yellow line shows the random and time-varying response a source of errors in the test.
This yellow line is also for troubleshooting afterward.

The tool logs the experimenter's actions with time stamps.
The ``\texttt{MEMO5s}'' button records a voice memo for five seconds and saves the recording using a time-stamp-based unique name.
It is an excellent practice to take photos while conducting experiments using smartphones.
The timers of smartphones and computers share the synchronized network time using NTP (Network Time Protocol)\cite{ntp1991ieee}.
The synchronization accuracy is enough to align logged events, voice memos, and smartphone photos.

\subsubsection{Supporting tools}
\begin{figure*}[tbp]
\begin{center}
\includegraphics[width=\hsize]{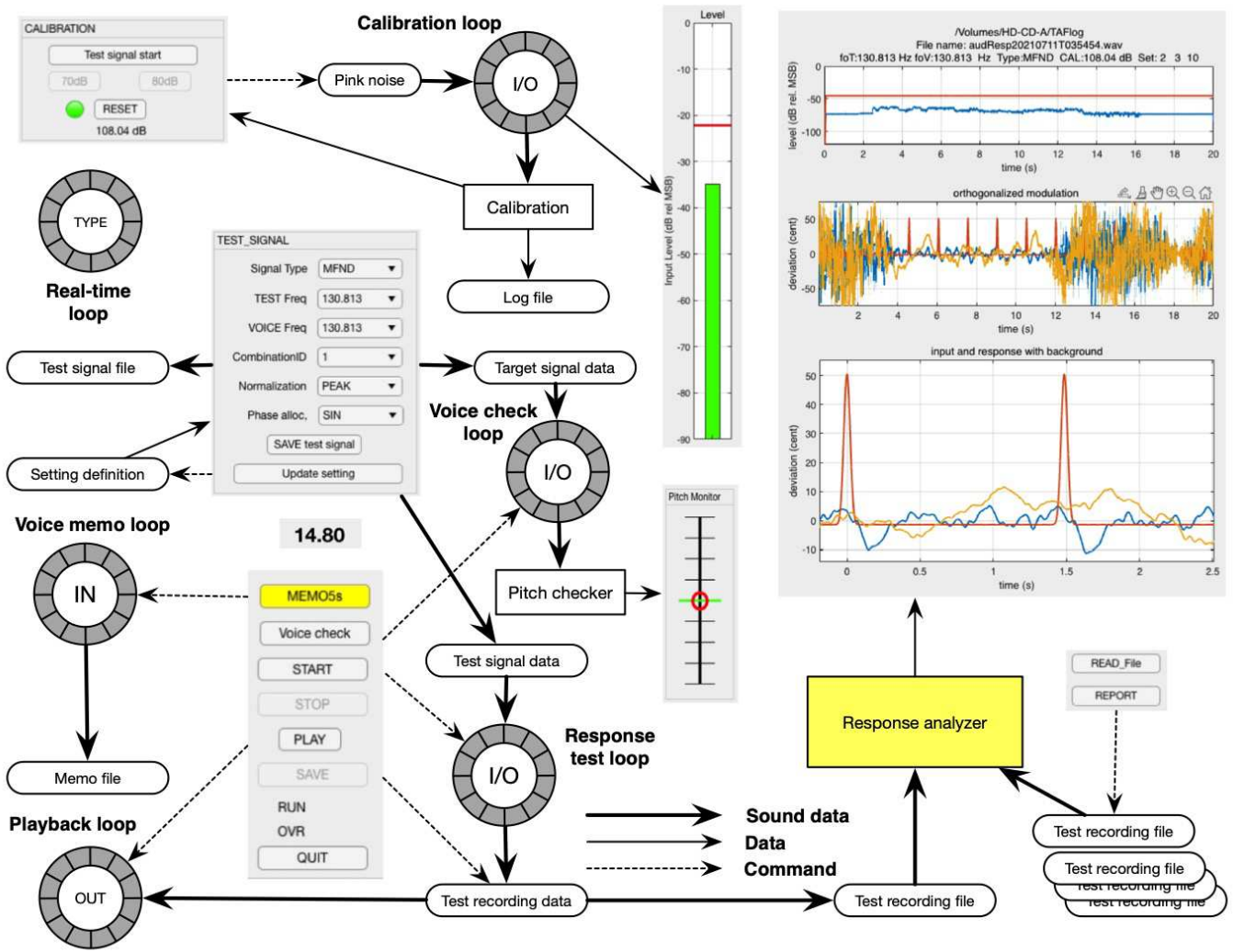}
\caption{Functional diagram of the interactive tool.}
\label{fig:toolStructure}
\end{center}
\end{figure*}
The GUI tool calls MATLAB functions to generate test and target sounds to analyze the test results.
These functions provide offline tools for detailed analyses and generating reports.
We developed an acoustic measurement tool\cite{kawahara2021ISTool} by adopting CAPRICEP\cite{kawahara2021icassp} to the simultaneous multi-attributes measurement method\cite{kawahara2020simultaneous}.
The tool assesses the recording conditions and sound stimuli presentation referring to the recommendations\cite{Rita2018ajsp,svec2010ajsp}.

\section{Design and implementation}
We designed the tool for assisting the exploratory stage of investigations where flexibility is more important than precision.
It is also crucial to record the history of trials with settings and data and avoid retrospective editing.
Because this tool encourages trial and error in the exploratory stage, we do not introduce strict security mechanisms.
Instead, we time-stamped crucial actions and data.
As mentioned before, saving data before displaying analysis results helps prevent (hopefully unintentional) cherry-picking.

Figure~\ref{fig:toolStructure} shows functional structure of the interactive tool.
It shows the relations between GUI parts and functions, data, files, and displays.
We placed technical details in appendices because they may disturb readers to acquire appropriate mental models.
In this section, we focus on providing a general understanding.

\subsection{Vocal response measurement}
Gray rings represent real-time processing loops.
The body of the experiment uses the \textsf{\textbf{Response test loop}} that is in the bottom center.
The ``\texttt{SAVE}'' button starts this loop and playback data from \textsf{\textbf{Test signal data}} and simultaneously acquiring the participant's voice.
This loop updates only the time display while the loop is running.
This minimum update is to prevent data loss caused by glitches in the real-time processing.

The \textsf{\textbf{Test recording data}} holds the acquired data (L-channel: voice, and R-channel: loop-back signal).
The \textsf{\textbf{Playback loop}} is to check recording errors, if any.
The ``\texttt{SAVE}'' button issues a command to save the acquired data to a uniquely-named file.
The command also starts the analysis of the saved file.

The \textsf{\textbf{Response analyzer}} is a MATLAB m-function.
The experimenter can use the function for offline processing programming.

\subsection{Voice check}
The essential secondary loop is the \textsf{\textbf{Voice check loop}}.
It playback the data in the \textsf{\textbf{Target signal data}} and simultaneously acquires the participant's voice.
It calculates $f_\mathrm{o}$ in real-time and updates the pitch display.
It also updates the time display.
Please refer to technical details in designing the real-time pitch information display.

\subsection{Calibration}
The \textsf{\textbf{Calibration loop}} also uses simultaneous playback and recording.
It updates the level display that displays the RMS value and the peak value.
It also updates the time display.
The vertical axis represents the level in terms of the full scale of the input signal.

\subsection{Voice memo}
The voice memo uses a real-time input loop, the \textsf{\textbf{Voice memo loop}}.
It updates the time display because it only records five seconds.

\section{Conclusions}
We introduced a practical procedure for measuring the involuntary response of voice $f_\mathrm{o}$ to auditory stimulation with frequency-modulated $f_\mathrm{o}$.
The procedure uses an interactive and real-time tool designed for investigating relevant settings of substantial experiments.
The subjects' task is to produce a sustained vowel keeping pitch constant while listening to a test sound.
The presentation of the test sound has two modes, and one mode uses a headphone the other mode uses a loudspeaker.
Each test session lasts about 20 seconds, and the analysis result shows up in several seconds.
We designed and implemented the tool to facilitate the exploratory phase of investigations.
This article focuses on providing an appropriate mental model of the tool's behavior to experimenters.
We provided technical details in appendices.
The tools described in this article and related materials are accessible in the first author's GitHub repository\cite{kawahara2020gitHk}.

\section*{Acknowledgment}
This work was supported by JSPS (Japan Society for the Promotion of Science) Grants-in-Aid for Scientific Research Grant Numbers JP18K00147, JP18K10708, JP19K21618, and JP21H04900.
We also thank Liao Jiahui, a master student of Toyohashi University of Technology for comments on defects and suggestions on improvement of the tools.

\bibliographystyle{IEEEtran}

\bibliography{kawaharaAudSt.bib}

\appendix

\section*{MATLAB implementation}

We implemented these tools using MATLAB and three toolboxes, signal processing, dsp, and audio\cite{rtAudioMatlab,rtAudioPerfMatlab,rtAudioPerfTestMatlab}.
The audio toolbox enables stream processing at an audio sampling rate by providing the system object of MATLAB.
The system object interfaces the driver of the audio device and the procedure written in MATLAB.

The audio toolbox provides three types of system objects, \texttt{deviceReader}, \texttt{deviceWriter}, and \texttt{playRec} for input stream (for example, microphone input) processing, output stream (for example, sound output for loudspeaker), and synchronized input and output stream processing, respectively.
Initially, we used \texttt{playRec} for implementing all real-time procedures in the tools.
The toolbox assigns different device names for different system objects, even though the connected hardware is the same. 

\subsection{Real-time loop}
The synchronized input and output stream processing has the following structure.
{\footnotesize
\begin{verbatim}
playRec = audioPlayerRecorder(<Setting>); 
while <condition> 
    <User procedure-1: prepare audioToDevice> 
    audioFromDevice = playRec(audioToDevice); 
    <User procedure-2: Use audioFromDevice>
end
\end{verbatim}
}
The first line assigns the device that is capable of synchronized input and output.
The while loop is the body of the real-time process.
The object reads and writes repeatedly at a constant speed.
Therefore, the total processing time of the user procedures needs smaller than the repetition interval.
We made all signal generation procedures operate outside this real-time loop and store the generated signals to shared buffers.
The user procedure-1 in the loop reads the data for \texttt{audioToDevice} from this shared buffer.

This processing structure makes real-time programming very easy.
The system object conceals details of real-time processes from MATLAB programmers.

\subsection{Issues with \texttt{App Designer}}
\texttt{App Designer} is an interactive development environment for designing an app layout and programming its behavior\cite{appDesigner}. 
It makes GUI application design easy.
However, interactivity support and graph rendering are not compatible with real-time processing because the former introduces many interruptions, and the latter requires significant computing resources.

We disabled interactivity of each graph using the ``\texttt{disableDefaultInteractivity}'' function.
We set ``\texttt{limitrate}'' option to the graph rendering function ``\texttt{drawnow}.''
This option allows skipping update rendering when the total processing time exceeds the repetition interval of the real-time loop.

\subsection{Issues in \textbf{\textsf{Response test loop}}}
We tested the real-time latency and glitches (overrun and underrun) of macOS and Windows 10.
We found that simultaneous input and output using \texttt{audioPlayerRecorder} causes underrun in-frequently.
We decided to run \texttt{audioDeviceWriter} and \texttt{audioDeviceReader} simultaneously and let the audio interface to synchronize the input and output.

\subsection{Real-time pitch monitor}
The buffer length used by all real-time loops of the tool is 1024, and the sampling frequency is 44100~Hz.
This buffer length is too short for calculating $f_\mathrm{o}$ of low-pitched male voices.
We implemented a ring buffer that has appropriate buffer length and updated the data in the \textbf{\textsf{Voice check loop}}.
We calculated two short-time Fourier transforms of two time-windowed segments (one sample shifted segment pair) and calculated $f_\mathrm{o}$ using the instantaneous frequency.

We applied a first-order IIR low-pass filtering to the raw $f_\mathrm{o}$ value to calculate the value for the pitch monitor display.
This low-pass filtering makes the display move smoothly and is appropriate for instructing participants.

\subsection{Calibration}
We also implemented a ring buffer for calcularing the RMS value for smoother level display.
The \textbf{\textsf{Calibration loop}} updates the ring buffer.

\end{document}